\documentstyle[prl,aps]{revtex}
\begin{document}
%\draft
\def\be{\begin{equation}}
\def\ee{\end{equation}}
\def\ra{\rightarrow}
\def\o{\over}
\def\l{\left}
\def\r{\right}
\title
{Kruskal Dynamics For Radial Geodesics. I}

\author{Abhas Mitra}
\address{Theoretical Physics Division, Bhabha Atomic Research Center,\\
Mumbai-400085, India\\ E-mail: amitra@apsara.barc.ernet.in}

%\date{\today}

\maketitle

\begin{abstract}
The total spacetime manifold for a Schwarzschild black hole (BH)  is
described by the Kruskal coordinates $u=u(r,t)$ and $v=v(r,t)$, where $r$
and $t$ are the conventional Schwarzschild radial  and time coordinates
respectively.  The relationship between  $r$ and $t$ for a test particle
moving on a radial or non-radial geodesic is well known. Similarly, the
expression for the vacuum Schwarzschild derivative for a  geodesic, in
terms of the constants of motion, is well known.  However, the same is not
true for the Kruskal coordinates; and, we derive here the expression for
the Kruskal derivative for a radial geodesic in terms of the constants of
motion. In particular, it is seen that the value of $\mid du/dv\mid (= 1$)
is regular on the Event Horizon of the Black Hole.  The {\em regular
nature} of the Kruskal derivative is in sharp contrast with the
Schwarzschild derivative, $\mid dt/dr\mid =\infty$, at the Event Horizon.
We also explicitly obtain the value of the Kruskal coordinates on the
Event Horizon as a function of the constant of motion for a test particle
on a radial geodesic.
\end{abstract}

\newpage
%\vskip 1cm
\section{Introduction}
It is known for more than 80 years that
the
region exterior to a point mass or the event horizon ($r >r_g=2m$) of a Schwarzschild
Black Hole (BH) can be described by the
vacuum Schwarzschild metric\cite{1,2}:
\begin{equation}
ds^2 = g_{tt} dt^2 +g_{rr} dr^2 + g_{\theta \theta} d\theta^2 + g_{\phi \phi} d\phi^2
\end{equation}
where
$g_{tt} =(1- 2m/r)$, $ g_{rr} =-(1-2m/r)^{-1}$, $ g_{\theta
\theta}=-r^2$, and $g_{\phi \phi} = -r^2 \sin^2 \theta$.
 Here, we are working with a spacetime signature of +1, -1, -1, -1 and
$r$ has a distinct physical significance as the {\em invariant
circumference radius}.  The coordinate time $t$ too has a physical
significance as the proper time of a distant inertial observer $S_\infty$.
At $r=2m$, $g_{rr}$
blows up and as $r <2m$, the $g_{tt}$ and $g_{rr}$ suddenly exchange their
signatures  though the signatures of $g_{\theta \theta}$ and $g_{\phi
\phi}$ remain unchanged. The detail dynamics of a ``test particle'' in the
vacuum external spacetime is well known for a very long time and
discussion on it is contained in practically every text book or monograpgh
on classical General Theory of Relativity (GTR). One of the key aspects for
studying the kinematics of a test particle is the knowledge about the
relevant derivative of the spatial coordinate with the temporal one. For
instance for any
geodesic having angular momentum or not, one knows the details about the
behaviour of the
Schwarzschild derivative $dr/dt$ or $dt/dr$. And the fact that $dt/dr
=-\infty$ blows up at the Event Horizon restricts the utility of the
Schwarzschild dynamics below the EH.
This tantamounts to the well known fact that the vacuum Schwarzschild
metric fails to describe the spacetime inside $r \le 2m$.

 On the other hand, we learnt in 1960 that both the
exterior and the interior regions of a BH may be described by 
a
one-piece coordinate system
suggested by Kruskal and Szekeres\cite{3,4}. Though, in the intervening 39 years
hundreds of articles have been written on Kruskal coordinates, and,
most of the treatises on GTR too regularrly deliberate upon the original
work of Kruskal and Szekers, the fact remains that sufficient effort has not
been made to study the kinematics of a test particle in terms of the
Kruskal coordinates so that one could have a better insight and
appreciation of the kinematics inside the EH. As a first step towards this
direction, in this paper, we would derive expressions for the Kruskal derivative
$du/dv$ for a radial geodesic. For the sake of completeness, we shall
start from the usual description about the Kruskal coordinates and first
derive the exact expression for the value of the Kruskal coordinates on
the EH ($u_H$ and $v_H$) in terms of $r$ and $t$. We shall show here 
that $u_H$ and $v_H$ are always non-zero and finite in general. More
importantly, we shall explicitly show that unlike the Schwarzschild
derivative, the Kruskal derivates are regular on the EH 
in accordance with the  singularity free nature of the Kruskal coordinates. 

\section{Kruskal Coordinates}
For the  region exterior to the EH (Sectors I \& III), the Kruskal
coordinates are defined as follows:
\begin{equation}
u=f_1(r) \cosh
{t\over 4m}; \qquad v=f_1(r) \sinh
{t\over 4m}; ~~r\ge 2m
\end{equation}
where
\begin{equation}
f_1(r) =\pm \left({r\over 2m} -1\right)^{1/2} e^{r/4m}
\end{equation}
Here the plus sign corresponds to ``our universe'' while the negative sign
corresponds to the ``other universe''\cite{1,2}. The ``other universe'' is
a legitimate mathematical solution of the Schwarzschild problem (irrespective of its observational
reality), and is a time reversed mirror image of ``our universe''.

And for the region interior to the horizon (Sectors II \& IV), we have
\begin{equation}
u=f_2(r) \sinh
{t\over 4m};\qquad v=f_2(r) \cosh
{t\over 4m}; ~~r\le 2m
\end{equation}
where
\begin{equation}
f_2(r) =\pm \left(1- {r\over 2m}\right)^{1/2} e^{r/4m}
\end{equation}

In terms of $u$ and $v$, the metric for the entire spacetime is
\begin{equation}
ds^2 = {32 M^3\over  r} e^{-r/2m} (dv^2 -d u^2) - r^2 (d\theta^2
+d\phi^2 \sin^2 \theta)
\end{equation}
The metric coefficients are regular everywhere except at the
intrinsic singularity $r=0$, as is expected. Since afterall the Kruskal coordinates
are defined using $r$ and $t$, for a proper understanding of the Kruskal dynamics,
it is necessary to recall the inter-relationship between the Schwarzschild
coordinates for a geodesic.

\subsection{Inter Relation Between Schwarzschild Coordinates}
For a test particle on a radial geodesic, the angular momentum is zero,
and there is only one conserved quantity, the energy of the particle (per
unit rest mass), $E$,
 as measured by a distant inertial observer:
 
 \begin{equation}
 E \equiv{dt\over ds} (1-2m/r)
 \end{equation}
 where $s$ is the proper time. For a massless particle like a photon,
we have $E=\infty$, otherwise $E$ is finite. For a radial geodesic, the
motion of the particle is determined by (see Chandrasekhar, pp. 98)\cite{5}
\begin{equation}
{dr\over ds} = -\sqrt{E^2 -(1-2m/r)}
\end{equation}
and
\begin{equation}
{dt\over ds} = {E\over 1-2m/r}
\end{equation}
so that 
\begin{equation}
{dt\over dr} = - {E (1-2m/r)^{-1}\o \sqrt{E^2 -(1-2m/r)}} 
\end{equation}
Clearly as $r\rightarrow 2m$, $dt/dr \rightarrow -\infty$. 
 On the other
hand, we do not expect such irregular behaviour for the Kruskal derivative.

Here note that if the particle is released from rest ($dr/ds=0$) at $r=r_i$ at
$t=0$, from Eq. (8), it is seen that \cite{5}
\be
E^2 = (1-2m/r_i)
\ee
or,
\be
r_i/2m = (1-E^2)^{-1}
\ee
It is convenient to introduce a (cylic) parameter $\eta$ through
\begin{equation}
r ={r_i\over 2} (1+\cos \eta) = {2m\o 1-E^2} \cos^2 (\eta/2) = r_i \cos^2 (\eta/2)
\end{equation}
Obviously, $\eta=0$ when $r=r_i$ and at the EH, we have
\be
\eta=\eta_H = 2 \arcsin E; \qquad r=2m
\ee
Now after some manipulation, Chandrasekhar arrived at the following Eq. involving
$t$ and $\eta$\cite{5}:
\be
{dt\o d\eta} = E \l({r_i\o 2m}\r)^{1/2}~ {\cos^4 (\eta/2)\o \cos^2
(\eta/2) - \cos^2 (\eta_H/2)}
\ee
This Eq. can be integrated to find the exact relation between $t$ and $r$
for a radial geodesic (actually, even for non-radial geodesic this Eq.
would hold good):
\be
{t\o 2m} = E \l({r_i\o 2m}\r)^{3/2} [{1\o 2} (\eta +\sin \eta) + (1-E^2) \eta]
+  \ln\l[{\tan (\eta_H/2) + \tan (\eta/2) \o \tan (\eta_H/2) - \tan (\eta/2)}\r]
\ee
The above Eq. may also be written without introducing $\eta_H$ and $E$ explicitly:
 (see pp. 824 of ref.[1] or pp. 343 of ref.[2]):
\begin{equation}
{t\over 2m} = \ln\mid{(r_i/2m-1)^{1/2} + (r_i/r -1)^{1/2} \over (r_i
/2m-1)^{1/2} - (r_i/r -1)^{1/2}}\mid + \left({r_i\over 2m}-1\right)^{1/2}
\left[\eta + \left({r_i\over 4m}\right)(\eta +\sin \eta)\right]
\end{equation}

We find from Eqs. (16-17) that, as $r\rightarrow 2m$ from Sector I, the logarithmic
term blows up and $t\rightarrow \infty$, which is a well known
result. Further Kruskal coordinates envisage that approach to the EH from
the Sectors III \& IV corresponds to $t=-\infty$.

\subsection{Kruskal Coordinates on the Event Horizon}
In Sectors I \& III, Kruskal coordinates obey the relation
\be
{u\o v} = \coth{t\o 4m}
\ee
And since $r\ra 2m$ corresponds to $t\ra \pm\infty$, at the EH, we have
\be
{u_H\o v_H} =\pm 1;\qquad r=2m
\ee
On the other hand,  in Sectors II and IV, we see
\be
{u\o v} = \tanh{t\o 4m}
\ee
and as
 $r\ra 2m$,  $t\ra \pm\infty$, we are led to the same Eq. (19).
In the same limit, $r\ra 2m$ and $t\ra \pm \infty$, we find that
\be
u_H^2 =v_H^2 \ra f_1^2 \exp{t\o 2m}
\ee
It might appear that since $f_1(2m)=f_2(2m) =0$ on the EH,  we would have
$u_H=\pm v_H =0$.
 But this is incorrect because the temporal part of $u$
and $v$ tend to blow up much more rapidly on the EH. And one has to
carefully obtain the actual values of $u_H$ and $v_H$ by working out
appropriate limits

To do so we introduce a new variable 
\be
z = r_i /2m -1 = {E^2\o 1-E^2}
\ee
and, let, in the vicinity of the EH,
\be
r/2m =1+\epsilon;\qquad \epsilon\ra 0
\ee
so that
\be
f_i^2(2m) \ra \epsilon
\ee
Then, in the vicinity of the EH,  by retaining terms first order in
$\epsilon$, we can rewrite Eq.(17) as
\begin{equation}
{t\over 2m} = \ln \mid {z^{1/2} + z^{1/2} \left(1 -{\epsilon r_i\over 4 mz}\right)\over
z^{1/2} - z^{1/2} \left(1 -{\epsilon r_i\over 4 mz
}\right)}\mid + \left({r_i \over 2m}
-1\right)^{1/2} \left[\eta + \left({r_i\over 4m}\right) (\eta + \sin \eta)\right]
\end{equation}
As $\epsilon \rightarrow 0$, the logarithmic term in the above expression becomes
\begin{equation}
A(t) = \ln \mid {1 + 1 -{\epsilon r_i\over 4 mz}\over
1 - 1 +{\epsilon r_i\over 4 mz}
}\mid \ra  \ln{8mz \over r_i \epsilon}
\end{equation}

Then, using Eqs. (24) and (26), we find
\begin{equation}
f_1^2 \exp{(A)} \rightarrow {8 e m z\over r_i} = 4e (1 - 2m/r_i) = 4e E^2
\end{equation}

Now considering the other terms in the expression for $t/2m$ in Eq.(25), we find
that, in this limit,

\begin{equation}
u_H^2 =v_H^2 = 4e(1- 2m/r_i) \exp\left\{ \left({r_i\over
2m}-1\right)^{1/2}\left[\eta_H +\l({r_i\o 4m}\r) (\eta_H
+\sin{\eta_H})\right]\right\}
\end{equation}

In terms of $E$, we have
\begin{equation}
u_H^2 =v_H^2 = 4e E^2 \exp~E \left[\eta_H + {\eta_H
 +\sin{\eta_H}\o 2 \sqrt{1-E^2}}\right]
\end{equation}

One would have $u_H=v_H=0$ if $E=0$ or, if the test particle is injected
{\em from rest} right at the EH. Clearly, this is unphysical, and
thus we  see that  $u_H$ and $v_H$ are non-zero. Further,
 for a finite value
of $r_i/2m$ or for $E<1$, they are finite too. 
The finiteness of $u$ and $v$ at the EH is physically appealing because
$u$ and $v$ are expected to be completely regular at the EH.
However for $r_i/2m =\infty$ or $E=1$, we find $u_H^2= v_H^2 = \infty$. 

On the other hand, since $r=r_i$ at $t=0$, by using the definition of $u$
and $v$, we find that the initial values of
\begin{equation}
u^2=u_i^2 = (r_i/2m -1) = {E^2\o 1-E^2}
\end{equation}
and
\begin{equation}
v^2 =v_i^2 =0
\end{equation}

\section{Kruskal Derivative: A Direct Approach}
Having shown that $u_H$ and $v_H$ are, in general, non-zero, 
we are now in a position to evaluate the Kruskal derivative,
the key ingredient for studying the Kruskal dynamics for a radial geodesic.
We first confine ourselves to Sector I.
By differentiating $f_1(r)$ (Eq.[3]) with $r$ we obtain
\begin{equation}
{df_1\over dr} = {\pm r\over 2m}~ {e^{r/4m}\over 4m}~ (r/2m-1)^{-1/2}
\end{equation}
Then by directly differentiating Eq.(2) by $r$, we find that irrespective
of the sign of $df_1/dr$, we will have
\be
{du\o dr} = {df_1\o dr} \cosh {t\o 4m} + {f_1\o 4m} \sinh{t\o 4m} ~{dt\o dr}
\ee
Interestingly, in all the sectors, we obtain the same functional form of $du/dr$.
Using Eqs.(2) and (4) in the foregoing Eq., we see that
\be
{du\o dr} = {u\o 4m} (1-2m/r)^{-1} + {v\o 4m}  ~{dt\o dr}
\ee
On the other hand by differentiating Eqs. (4) and (5), we find that

\begin{equation}
{df_2\over dr} = {\mp r\over 2m}~ {e^{r/4m}\over 4m}~ (r/2m-1)^{-1/2}
\end{equation}
and
\be
{du\o dr} = {df_2\o dr} \sinh {t\o 4m} + {f_2\o 4m} \cosh{t\o 4m} ~{dt\o dr}
\ee
And by using Eqs. (4) and (35) into the foregoing Eq., we obtain the {\em same
expression} (34) for $du/dr$ in Sectors II \& IV.
Further, using Eq.(10) in (34), we obtain the ultimate expression for
\be
{du\o dr} = {(1-2m/r)^{-1}\o 4m} \left[ u - {vE\over \sqrt {E^2 -1 +2m/r}}\right]
\ee
valid in all the sectors.
Similarly,  we obtain the ultimate functional form of $dv/dr$ which is
valid for {\em all the sectors}:
\be
{dv\o dr} = {(1-2m/r)^{-1}\o 4m} \left[ v - {uE\over \sqrt{E^2 -1 +2m/r}}\right]
\ee

And, the general value of $du/dv$ in any Sector is obtained
 by dividing Eq.(37) with (38):
\be
{du\o dv} = {  u - {vE\over \sqrt{E^2 -1 +2m/r}}\o
 v - {uE\over \sqrt{E^2 -1 +2m/r}}}
\ee

\subsection{Kruskal Derivative at the Event Horizon}
Since $u$ and $v$ are expected to be differentiable smooth continuous
(singularity free) functions everywhere except at $r=0$, and also since
the ``other universe'' is a mirror image of ``our universe'', we expect that
the value of $du/dv$ for any given $r$ must be the same, except for a
probable difference in the signature, in both the universes.
The meaningful way to find the value of $du/dr$ at the EH will be to concentrate
on the Sectors II \& IV for which $u_H =-v_H$

\be
{du\o dv} \ra {  u - v\o
 v - u} = {2 u_H \o 2 v_H} = -1; \qquad r= 2m
\ee
The Eq. (39) for the Kruskal redivative, however, tends to yield a ``$0/0$''
form at $r=2m$ for Sectors I \& III having $u_H= v_H$. But as mentioned above,
 we expect this  $0/0$ form to acquire the
value $du/dv = +1$ because these Scetors are the mirror images of Sectors
II \& IV. Otherwise the whole idea of having an extended time symmetric Schwarzschild manifold 
 would be inconsistent. Thus, in
general, we must have

\be
{du/dv} = \pm 1; \qquad r=2m
\ee
The fact that we must have $du/dv =+1$ for the Sectors I \& III
 can be reconfirmed in the limiting case of 
 $u_H^2 =v_H^2 =\infty$ for $E=1$ or $u_H
=v_H =0$ for the (unphysical case) $E=0$ directly
by using L' Hospital's theorem.

Note that, by this rule, we can write,

\be
\lim_{u \ra 0(\infty), v\ra 0 (\infty)} {u\o v} = \lim_{u \ra 0 (\infty),
v\ra 0 (\infty)} {du/dr\o dv/dr} 
\ee

In any case, from Eq. (19), we already know that $u/v =\pm 1$ at
$r=2m$. Then we can rearrange the foregoing Eq. as

\be
 \lim_{u \ra 0 (\infty), v\ra 0 (\infty)} {du/dr\o dv/dr} =\pm 1
\ee
or,
\be
 \lim_{r\ra 2m} {du\o dv} =\pm 1
\ee

\section{A Different Route}
It may be of some interest to rederive the limiting value of $du/dv$ by
using other generic relationships between $u$ and $v$. As before, to avoid
$0/0$ forms, we work with Sectors III \& IV. In particular, in
Sector III, we have
\be
{u\o v} = \coth {t\o 4m}
\ee
By differentiating this equation w.r.t. $v$, we obtain
\be
{1\o v} {du\o dv} -{u\o v^2} = -{1\o 4m} {1\o \sinh^2(t/4m)} ~
{dt\o dv}
\ee
By recalling that $\sinh(t/4m) = v/f_1$, we rewrite the above Eq. as
 \be
 {du\o dv} -{u\o v} = {-f_1^2\o 4m} {1\o v} {dt\o dv}
 \ee
Now, from Eqs. (10) and (38), note that
\be
{dt\o dv} = {dt\o dr} {dr\o dv} = - {4mE\o v\sqrt{E^2 -1 +2m/r} -uE}
\ee
And the limiting value of
 \be
 {dt\o dv} \ra{4m\o u-v} = {4m\o u_H +u_H} ={2m \o u_H} \qquad r\ra 2m 
 \ee
And since $f_1(2m) =0$, we find from Eq. (47) that

 \be
 {du\o dv} -{u\o v} =0; \qquad r=2m
 \ee

Or,
 \be
 {du\o dv} ={u\o v} = {u_H\o -u_H} =-1; \qquad r=2m ~(Sector~ III + IV)
 \ee

Similarly, for the sake of overall consistency, in Sectors, I \& III, we
must have $du/dv =+1$  at $r=2m$.

\section {A Different Consideration}
Actually we could have obtained the above derived unique result in a
relatively simpler manner by 
 differentiating the Global Eq.
 \be
 u^2 -v^2 = (r/2m -1) e^{r/2m}
 \ee
   w.r.t. $v$ :
\be
2u~ {du\o dv} - 2 v= {r\o 4m^2} e^{r/2m} ~ {dr\o dv}
\ee
First let us note from Eq.(37) that in Sectors II \& IV, the limiting
value of

\be
 {dr\o dv} \ra {4m (1-2m/r)\o v-u} \ra  {2m (1-2m/r) \o v_H} =0;\qquad r\ra 2m
\ee
Then, by using this above Eq. in (53), we find

\be
u_H~ {du\o dv}  = v_H; \qquad r\ra 2m
\ee
so that
\be
 {du\o dv}  = {v_H\o -v_H} = -1; \qquad r\ra 2m
\ee
 
 \section{Conclusions}
The Kruskal coordinates were found way back in 1960, and
in the present paper, we have worked out some aspects of the kinematics of
a test particle following a radial Kruskal geodesic. To attain this we
used, for the first time, the precise value of $u_H$ and $v_H$
as a function of the initial conditions of the problem $r_i$,  $m$ or $E$.
 It
is clearly found that $u_H^2=v_H^2$ is non-zero in general.

We then proceeded to obtain the expression for the Kruskal derivative in
terms of $m$, $E$ and $r$. We found that the {\em Kruskal derivative is regular at
the EH unlike the Scharzschild deivative(s)} where $dt/dr = -\infty$ at the EH.

In particular $du/dv =+1$ at the EH if we  consider the ``other
universe''whose existence is
suggested by the full Kruskal manifold, and which is a time reversed
version of ``our universe''. But, if we move to the ``our universe'',
 the expected value of $du/dv =1$ at the EH.

 The {\em regular nature of
the Kruskal derivative is in keeping with the notion that Kruskal
coorinates are free of singularities at the EH}.

 In a subsequent paper, we
shall find out other important features of the Kruskal dynamics vis-a-vis
the well known Schwarzschild dynamics.

\section{Acknowledgements}
The author wishes to thank Profs. P.C. Vaidya and P.S. Joshi for some
relevant useful discussions.

\end{document}